# Growth of ZnO nanolayers inside the capillaries of photonic crystal fibres

I. Konidakis*, M. Androulidaki, G. Zito, S. Pissadakis

*Foundation for Research and Technology-Hellas (FORTH), Institute of Electronic Structure and Laser (IESL), P.O. Box 1315, 71 110, Heraklion, Greece*

*Corresponding author:

ikonid@iesl.forth.gr, Tel: +30-2810-391835, Fax: +30-2810-391318.

**Abstract**

In this study, we describe the formation of ZnO nanolayers inside the air capillaries of a silica photonic crystal fibre (PCF), targeting random laser and organic vapor sensing applications. ZnO nanolayers were developed by infiltrating the capillaries of the silica PCF with Zn-acetate/methanol solutions of various concentrations, followed by annealing treatments. The growth and morphology of the synthesized ZnO nanolayers were characterized by means of scanning electron microscopy (SEM), and found to be affected by the concentration of the Zn-acetate/methanol infiltration solution. For low concentrations inspection with SEM revealed the formation of 25 and 100 nm thick ZnO nanolayers across the entire length of the infiltrated capillaries, whereas increasing the Zn-acetate concentration resulted to the formation of randomly placed isolated ZnO nanorods. Room temperature photoluminescence spectra of the ZnO nanolayers inside the PCF were measured and compared with the corresponding spectra reported for ZnO structures formed on typical surfaces.

1. Introduction

Over the last two decades zinc oxide (ZnO) has attracted enormous scientific interest while intense research has been focused on the synthesis and characterization of novel ZnO nanostructures of various shapes and properties [1-4]. The reason behind this is the extensive use of such nanostructures on a wide variety of sensing, solid state lasers, piezoelectronic and optoelectronic applications. In particular, the wide direct bandgap of ZnO (~3.3 eV) and its large exciton binding energy (60 meV) at room temperature make it promising as a solid state laser material [5,6]. On the same time the high chemical stability, amenability to doping, and relatively low cost of ZnO crystalline nanostructures render them attractive materials for being implemented in various types of gas and humidity sensing devices [7-12]. The sensing processes are mainly based on the alteration of ZnO electronic and optical properties upon its oxidation via the absorption of gas molecules, as for instance molecules of volatile organic solvents such as methanol, ethanol and hexane [7,8], carbon monoxide [9], hydrogen gas [10], and ozone [11], on the surface of the ZnO nanostructures. Furthermore, the biocompatibility and non-toxicity of ZnO allows its safe usage in biosensors and biofunctional applications [13].

In the present study we demonstrate for the first time the formation of ZnO crystalline nanolayers inside the air capillaries of a solid core, all-silica photonic crystal fibre (PCF), targeting potential random laser and gas- and bio-sensing applications. A typical PCF consists of a core and a periodic microstructure of air capillaries running along the entire length of the fibre. Light propagation through the fibre is confined by the periodic photonic crystal structure, making the fibre to act as a low-loss waveguide [14,15]. Upon infiltration of the PCF air capillaries with a high refractive index medium, the coupling of light between the core and the microstructured cladding of the fibre is altered, leading to measurable distinctive features in the optical spectra. In recent years, the infiltration of PCF capillaries with high refractive index materials such as polymers [16], ferrofluids [17], and glasses [18,19], has been highlighted for the realization of novel in-fibre devices suitable for sensing applications, magnetometers, and supercontinuum generation. On a similar manner nanoparticles attached on the internal capillary walls have been used to induce plasmonic cladding structures, and plasmonic tuned fibres for sensing applications are realized following such route [20]. Within this context, growing ZnO nanolayers inside the air capillaries of a PCF could create a potential fibre sensor device for many applications, i.e. including possible detection of external pressure changes if one considers the piezoelectric properties of the ZnO semiconductor.

The growth of ZnO crystalline nanolayers inside the air capillaries of a solid core silica PCF by means of pressure assisted infiltration methods was the main aim of the present study. It was achieved by adopting a 'solution base' growth technique previously described by Fernandez et al. for synthesizing ZnO nanostructured films on silica substrates via sol-gel route [7]. The developed ZnO-nanolayer/silica PCFs were characterized by means of scanning electron microscopy (SEM), while the photoluminescence spectra of the fibres are also considered.

## 2. Experimental

For the present work a commercially available air/silica solid core microstructured PCF was used (LMA-10 drawn by NKT Photonics Ltd.). The LMA-10 fibre has a periodic four-ring lattice of hollows channels with an average diameter of 2.85 μm and lattice constant of 6.4 μm as determined by SEM. Fibre samples were filled with Zn-acetate (Zn(CH$_3$COO)$_2$·2H$_2$O)/methanol solutions of various concentrations, prepared by adding the appropriate amount of Zn-acetate in methanol and stirring for 1 h at room temperature. The Zn-acetate concentration in methanol solutions employed was varied from 0.038 up to 0.6 M. Infiltration of fibre capillaries was done by means of pressure assisted infiltration technique (nitrogen gas used), while the process was monitored by an optical microscope to ensure that all fibre capillaries were filled with solution.

The one end-face of a ~15 cm long fibre was constrained into a chamber containing the solution and the air capillaries of the PCF were filled upon increase of the chamber pressure. Typical pressures used for the infiltration processes were of the order of magnitude of ~5 bar. After all fibre capillaries were filled with solution, the fibre was heated at 300 °C for 20 minutes. During this period the methanol solvent was evaporated totally out of the fibre capillaries as verified by optical microscopy. This infiltration and methanol removing process via annealing at 300 °C was repeated several times on the same fibre for growing thicker ZnO overlayers inside the PCF capillaries; i.e. repeating process depends on the concentration of Zn-acetate/methanol solution. Upon completion of the final infiltration, the fibre was annealed for another 2 h at the same temperature. Fig. 1 shows a schematic representation of the infiltration process, while Table 1 summarizes preparation parameters for the ZnO-nanolayer/LMA-10 fibres fabricated in the present study.

The formation and growth of ZnO nanolayers inside the PCF capillaries was examined by field emission scanning electron microscopy (JEOL, JSM-7000F). In order to inspect the nanolayers uniformity along the fibre length, samples were cleaved at various positions, and SEM scans were performed on all cross sections.

Room temperature photoluminescence (PL) measurements were performed by employing a He-Cd cw laser at 325 nm with full power of 35 mW. The PL spectra were measured using spectrometer with grating 600grooves/mm blazed at 300 nm, and a sensitive liquid nitrogen cooled CCD camera [21].

## 3. Results and Discussion

*3.1 Scanning electron microscopy (SEM)*

SEM shot of the cleaved end face of ZnO-nanolayer/LMA-10 fibre sample S1 is shown in Fig. 2a, while details of an infiltrated air capillary are shown in Fig. 2b. Fibre S1 was infiltrated eight times with Zn-acetate/methanol solution of 0.038 M. A first glance at Fig. 2a indicates that hardly any ZnO nanolayer has been formed within the air capillaries of the LMA-10 PCF. However a careful inspection (Fig. 2b) revealed that a thin layer of ZnO has been coated on the surface of the silica PCF capillaries. This nanolayer was formed within all capillaries of the fibre and it was found to have a typical thickness of ~25 nm. Moreover, SEM examination in different cross sections verified that the observed ZnO nanolayer was grown throughout the entire length of the infiltrated and thermally annealed region of the ZnO-nanolayer/LMA-10 fibre. However, from Fig. 2b the presence of regions where the ZnO nanalayers are thinner becomes apparent also. Such regions are located randomly across the length of the fibre capillary.

In an attempt to fabricate thicker ZnO nanolayers inside the capillaries, we initially tried to perform additional infiltrations with Zn-acetate/methanol solution of the same concentration, i.e. 0.038 M as reported in ref. [7] and used for fibre S1. However, despite increasing the number of infiltrations to ten and twelve times, the developed ZnO nanolayer was similar to that of S1 ZnO-nanolayer/LMA-10 fibre shown in Fig. 2b, i.e. S1 fibre prepared with eight infiltrations (see Table 1). Thus, the cumulative deposition process appeared to be ineffective in growing thicker ZnO PCF overlayers.

It was then decided to try higher concentrations of Zn-acetate/methanol solution for the fabrication of ZnO-nanolayer/LMA-10 fibres. Fibre sample S2 was prepared by following the same procedure as for sample S1 but with 0.152 M Zn-acetate/methanol solution instead of 0.038 M used before. Fig. 3a shows magnified area of a SEM shot of the cleaved end face of fibre S2, and 3b a focused shot on a fibre capillary. After considering the data of Fig. 3a it appears that the increase of Zn-acetate concentration augmented the thickness of the ZnO nanolayers inside the fibre capillaries. In particular, it can be estimated from Fig. 3b that a uniform layer of ~100 nm thickness is formed throughout the length of the capillaries. It was believed at that point, that there might be a direct correlation between the thickness of the ZnO nanolayer and the concentration of Zn-acetate in the methanol solution used for infiltration, i.e. bearing in mind that four times higher concentration resulted to approximately four times thicker ZnO nanolayers. In order to explore such possibility, fibres were infiltrated with Zn-acetate/methanol solutions of even higher concentrations, laying between 0.3 and 0.45 M. However, we found no difference in the thickness or properties of the ZnO nanolayers within these fibres when compared to that of fibre S2 shown in Fig. 3.

Remarkably though, noticeable differences on the morphology of the synthesized ZnO structures were observed only when the LMA-10 fibre was infiltrated with methanol solution of 0.6 M; at room temperature such concentration leads to an oversaturated solution of Zn-acetate in methanol. That is the highest concentration used for the fibres of the present study, i.e. sample S3. Fig. 4a shows magnified area of a SEM shot of a cleaved end face of fibre S3, while 4b-d present details of the ZnO nanolayers formed within various fibre capillaries. This specific concentration of Zn-acetate leads to the synthesis of ZnO nanorods inside fibre capillaries, instead of the ZnO nanolayers observed for fibres S1 and S2. However, SEM investigation reveals that such structures are only present inside some of the capillaries and they are located in random positions along the infiltration length of fibre S3, on the contrary to what was found for fibres S1 and S2 where uniform ZnO nanolayers were grown throughout the entire length of all capillaries.

The reason for such behavior is that upon formation of ZnO nanorods structures at a certain point of the fibre capillary, its diameter reduces significantly, and subsequently irregularities in the infiltration process are caused. As for instance, Fig. 4c shows that the formed ZnO nanorods almost block the fibre capillary, and thus, re-infiltration of the same capillary becomes almost impossible. Such behavior confines the formation of ZnO structure throughout the total length of the fibre capillaries. The aforementioned irregularities were observed also during the infiltration process, and because of that it was selected to infiltrate fibre S3 only twice, instead of eight times as for fibres S1 and S2 (Table 1). At this point, we note that it was also tried to achieve uniform re-infiltrations of fibre S3 by increasing the chamber pressure up to 10 bars instead of 5 bars used before. However, this attempt not only did not improve the re-infiltration process, but on the contrary it resulted to 'washing out' the ZnO nanostructures which have been formed from the initial two infiltrations. Future work will be focused on finding desired combination of solution and annealing parameters for achieving formation of uniform ZnO nanorods within the entire length of the fibre capillaries.

*3.2 Photoluminescence (PL) spectra*

Fig. 5 presents a typical photoluminescence (PL) spectrum of ZnO nanolayer grown inside the PCF capillaries, along with the corresponding spectrum of a ZnO film synthesized on glass substrate for the sake of comparison. We generally observe a good agreement between the two spectra, since both of them are dominated by a strong UV photoluminescence emission band at ca. 380 nm, which is the characteristic near band edge transition of the wide band gap of the ZnO semiconductor [4,22-29]. Remarkably though, the bandwidth of the PCF grown nanolayers is found to be significantly narrower, underlying the formation of better quality ZnO crystals inside the PCF capillaries. We believe that the confined capillary area facilitates better arrangement of surface related defects, and thus enhancing the crystalline quality of ZnO nanolayers.

In addition, the near band edge transition of ZnO nanolayers formed within fibre capillaries is blue-shifted compared to the corresponding band of structures formed on glass substrate, i.e. band shifts from to 384 to 379 nm corresponding to optical band gap energies of 3.22 to 3.27 eV respectively. Similar energy variations have been previously attributed to the size difference between ZnO grains, with higher values of band gap energies denoting smaller grain sizes [26,27]. Thus, in our case, it is possible that ZnO nanolayers formed on the open glass substrate exhibit larger grain sizes, in comparison to smaller ZnO grains synthesized within the constrict environment of the PCF capillaries. Complementary evidence for such proposal emerges from a pressure-dependent photoluminescence study of ZnO nanowires by Shan et al. [24], where the near band edge transition PL band shifts towards higher energy upon increasing pressure. In the case of ZnO grown inside the PCF capillaries, strain effects emerging from the lattice mismatch between the ZnO and the silica surface, as well as, differences in the thermal expansion coefficients of ZnO and silica may occur, leading to interfacial strain generation. In particular, the thermal expansion coefficient of bulk ZnO is reported ~$4\times10^{-6}$ K$^{-1}$ [30], while that of fused silica is approximately an order of magnitude smaller, namely $0.54\times10^{-6}$ K$^{-1}$ [31]. Due to the above effects, we believe that the silica capillaries exert compressive strain to the synthesized nanolayers leading to smaller ZnO grain sizes, and thus, inducing the noticeable blue shift of the band gap luminescence peak located at the 380 nm vicinity. The above assertion can be also supported by considering the data presented by Yamamoto et al. [32], where Young modulus values up to 300 GPa are reported for thin ZnO films, compared to the corresponding value of 73 GPa for silica glass.

Furthermore, the UV band of PL spectra of ZnO nanolayers within the PCF exhibits two distinct shoulders at ca. 360 and 420 nm, that are absent from the corresponding spectra of ZnO on glass substrate. On previous studies, assigned PL bands in the vicinity of 420 nm have been attributed to the existence ZnO-SiO$_2$ surface interactions and interface defects [22,23]. Inside the PCF silica capillary such interactions are enhanced due to the interior silica surface of the PCF capillaries where drawing induced defects exist, while giving rise to a shoulder at ca. 420 nm of the corresponding PL spectrum. Secondly, according to PL studies of ZnO single crystalline material, the lower wavelength shoulder at ca. 360 nm possibly originates from the free excitons and their first excited state in ZnO crystals [2,33].

Finally, noticeable differences are observed between the visible range of the two PL spectra shown in Fig. 5. In particular, the spectrum of ZnO nanolayer within the PCF capillaries exhibits a broad green-yellow emission profile, i.e. from 510 to 580 nm, while on the contrary such profile is almost absent from the corresponding spectra of ZnO film on glass substrate. Several studies attribute the existence of green-yellow emission profile to the presence of both zinc interstitials and oxygen vacancies within the synthesized nanolayers [4,25-29]. Thus it appears that ZnO nanolayers synthesized within fibres enclose some stoichiometric defects, while ZnO films synthesized on glass substrate are of lower defect concentration. We believe that such finding is due to the lack of oxygen within the fibre capillaries, when compared to normal ambient conditions of open surfaces. Nonetheless, such oxygen deficiencies can play important role in the exploitation of the ZnO overlayer in gas and humidity sensing experiments, exploiting fastly activated surface defects [8,12].

## 4. Conclusions

ZnO nanolayers were synthesized inside the capillaries of a silica PCF, by infiltrating the fibre with Zinc-acetate/methanol solution of various concentrations. Upon the concentration of the Zinc-acetate/methanol solution, different thickness and morphology overlayers were grown onto the PCF capillary walls. High concentration of Zinc-acetate/methanol solution was found to change the morphology of the synthesized ZnO structures from nanolayers to nanorods. Photoluminescence spectra of the ZnO nanolayers grown inside the fibre capillaries were found to be in good agreement with the corresponding spectra reported for ZnO structures formed on typical glass substrates, denoting growth of high quality ZnO crystalline domains within PCF capillaries.

**Acknowledgements**

The authors are grateful to A. Manousaki (IESL, FORTH) for her assistance with SEM studies, and to A. Klini and M. Konstantaki (IESL, FORTH) for useful discussions. This work was partially supported by the EU Project SP4-Capacities "IASIS" CN 232479 and the EU Project CA "ASPICE" CN 287637.

Note: G. Zito is now at Physics Dept., Università degli Studi di Napoli Federico II, via Cintia, Napoli, Italy.

**Table 1:** Summary of preparation parameters and resulted types of ZnO structures for the ZnO-nanolayer/LMA-10 fibres fabricated in the present study.

| Fibre sample | Zn-acetate/methanol concentration (M) | Number of infiltrations | ZnO structure type |
|---|---|---|---|
| S1 | 0.038 | 8 | nanolayer ~25 nm, within all PCF capillaries |
| S2 | 0.152 | 8 | nanolayer ~100 nm, within all PCF capillaries |
| S3 | 0.6 | 2 | various nanorods, within few PCF capillaries |

**Figure captions:**

**Fig.1:** Schematic representation of pressure assisted infiltration process of LMA-10 fibre with Zn-acetate/methanol solution.

**Fig. 2: (a)** SEM shot of the cleaved end face of infiltrated ZnO-nanolayer/LMA-10 fibre sample S1 (0.038 M Zn-acetate/methanol solution). **(b)** Detail of the ZnO nanolayer formed on the surface of a fibre capillary.

**Fig. 3: (a)** Magnified area of the cleaved end face of infiltrated ZnO-nanolayer/LMA-10 fibre sample S2 (0.152 M Zn-acetate/methanol solution). **(b)** Detail of the ZnO nanolayer formed on the surface of a fibre capillary.

**Fig. 4: (a)** Magnified area of the cleaved end face of infiltrated ZnO-nanolayer/LMA-10 fibre sample S3 (0.6 M Zn-acetate/methanol solution). **(b)** Detail of an infiltrated capillary from panel (a) where ZnO nanorods have been formed. **(c** and **d)** Formations of ZnO nanorods within S3 fibre capillaries from cross sections other than the one shown in panel (a).

**Fig. 5:** Room temperature photoluminescence (PL) spectra of ZnO nanolayers synthesized on glass substrate (black line) and inside the capillaries of a PCF (red line) presented on linear scale.

**Fig. 1:**

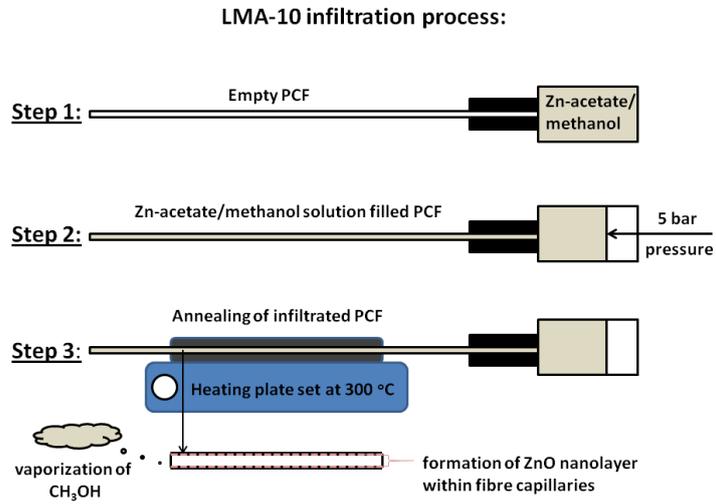

**Fig. 2:**

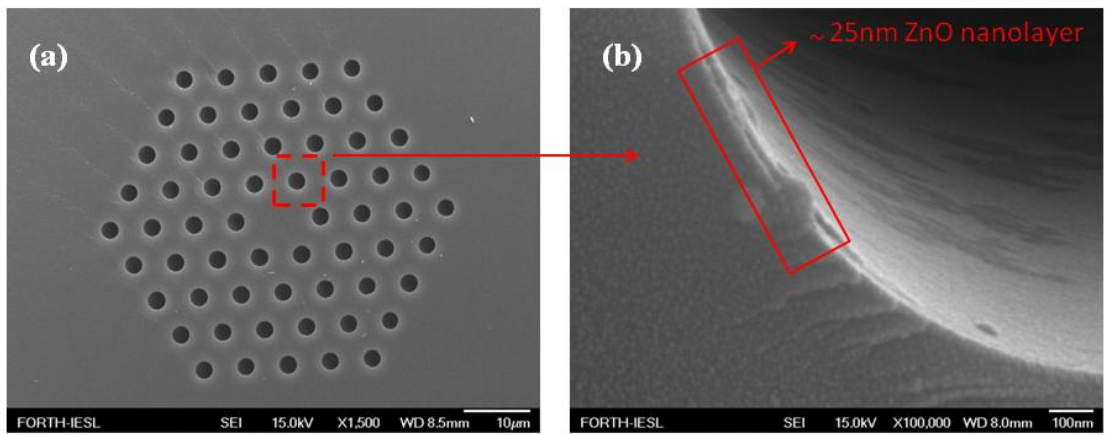

**Fig. 3:**

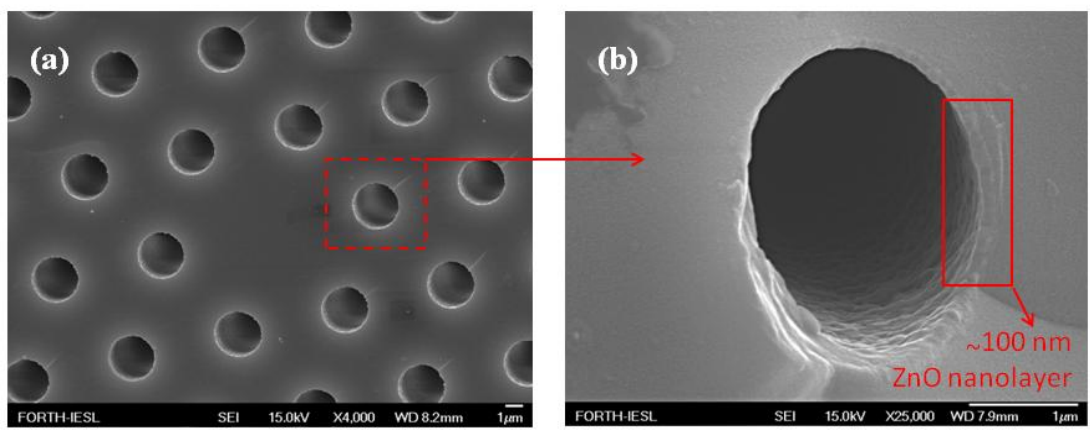

**Fig. 4:**

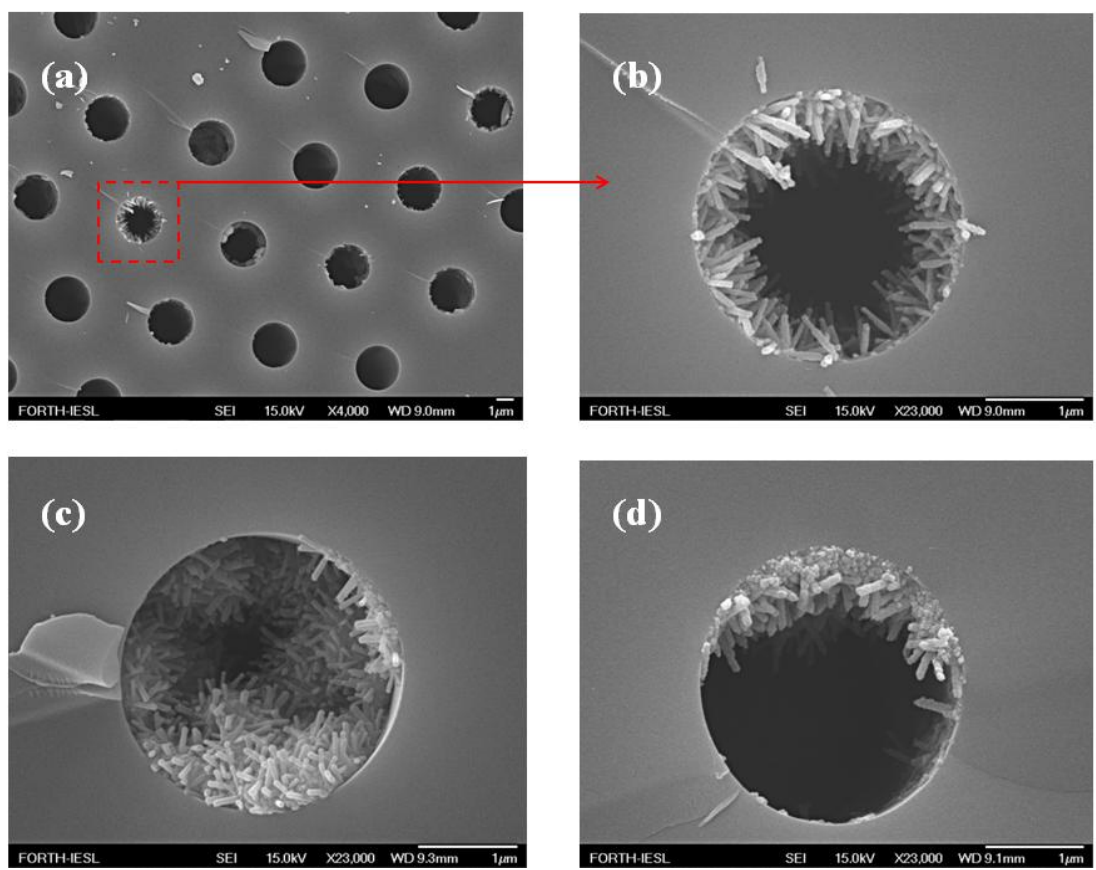

**Fig. 5:**

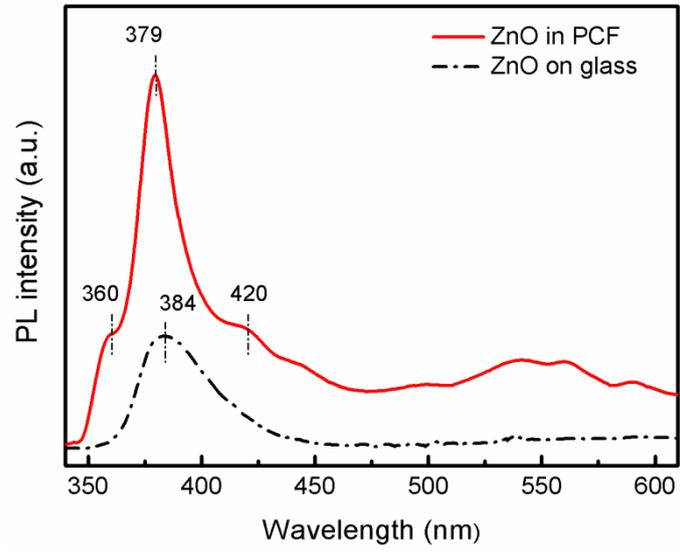